\documentclass[reprint,longbibliography,
 amsmath,amssymb,
 aps,pra
]{revtex4-1}

\usepackage{graphicx}
\usepackage{dcolumn}
\usepackage{bm}
\usepackage{lipsum}

\DeclareMathOperator{\tr}{tr}

\begin{document}

\preprint{APS/123-QED}

\title{Algebraic Probability-Theoretic Characterization of Quantum Correlations}

\author{Bin Yan}
 \affiliation{Department of Physics and Astronomy, Purdue University}
\date{\today}

\begin{abstract}
Quantum entanglement and nonlocality are inequivalent notions: There exist entangled states that nevertheless admit local-realistic interpretations. This paper studies a special class of local-hidden-variable theories, in which the linear structure of quantum measurement operators is preserved. It has been proven that a quantum state has such linear hidden-variable representations if, and only if, it is not entangled. Separable states are known to admit nonclassical correlations as well, which are captured by quantum discord and related measures. In the unified framework presented in this paper, zero-discordant states are characterized as the only states that admit fully consistent classical probability representations. Possible generalization of this framework to the quasiprobability representation of multipartite quantum states is also discussed.

%\begin{description}
%\end{description}

\end{abstract}

%\pacs{02.50.Cw, 03.65.Ud, 03.67.Mn}
                             
% 02.50.Cw	Probability theory
% 03.67.Mn	Entanglement measures, witnesses, and other characterizations
% 03.65.Ud	Entanglement and quantum nonlocality

\maketitle

\section{Introduction}
Composite quantum systems can exhibit correlations between subsystems that have no classical counterpart. Characterizing nonclassical correlations is of both fundamental and practical importance.
%Characterizing nonclassical correlations on one hand are of fundamental importance for the study of quantum physics; on the other hand, it helps to utilize the advantage of quantum resources over classical ones in practical tasks, especially in quantum information processing, such as quantum computations and communications.

Among various types of quantum correlations, entanglement is the most prominent. In the formalism of quantum mechanics, entangled states are formally defined as the states which can not be decomposed into a convex combination of product forms. This definition has a clear operational motivation: Entanglement can not be created by local operations and classical communication (LOCC) \cite{Bennett:1996,Horodecki:2009}. Another peculiarly nonclassical feature of compound quantum systems is nonlocality \cite{Brunner:2014}; e.g., there exist correlations between outcomes of separated measurements that can not be explained classically, in terms of local realistic theories. Such correlations are usually witnessed by violations of certain Bell inequalities \cite{Bell:1964,*Clauser:1969} satisfied by any local-hidden-variable (LHV) theories. A fundamental question which is natural to ask is  whether entanglement is merely a manifestation of nonlocality. Indeed, entangled states may lead to violations of suitable Bell inequalities. However, it has been known for some time that the notions of entanglement and nonlocality do not coincide: There exist entangled states that nevertheless admit LHV interpretations \cite{Werner:1989,Barrett:2002,Augusiak:2014,Bowles:2016}. Such states can not directly violate any Bell inequalities. Although it is known that these states do present some nonlocal features, termed hidden nonlocality \cite{Popescu:1995,Gisin:1996,Peres:1996,Masanes:2008}, if processed by certain operations prior to the Bell experiments, we are still far away from a complete understanding of the gap between entanglement and nonlocality.
%But these states do present some nonlocal features if processed by LOCC prior to the Bell experiments. This phenomenon, termed hidden nonlocality, can be revealed by various protocols, i.e., local filtering \cite{Popescu:1995,Gisin:1996}, tensor product several copies of the target state \cite{Peres:1996}, or joint process with an auxiliary state which does not violate Bell inequalities itself \cite{Masanes:2008}. Hence, it is believed that nonlocality in a sense characterizes quantum entanglement, although the nature of the existence of direct LHV models for certain entangled states still remains poorly understood.

On the other hand, the situation becomes more subtle in the realm of separable states. Though unentangled, separable states can exhibit nonclassical behaviors as well. These correlations are captured by quantum discord \cite{Henderson:2001,Ollivier:2001} and related measures \cite{Modi:2012}. It has been known that quantum discord also provides advantages in various quantum-information-processing tasks, e.g., deterministic quantum computation with one quantum bit (DQC1) \cite{Knill:1998,Datta:2008}, where quantum discord is created via the consumption of coherence of the subsystems \cite{ma:2016}.
%The most prominent example is the deterministic quantum computation with one quantum bit (DQC1) \cite{Knill:1998}, where quantum discord scales with the efficiency of the algorithm while entanglement is vanishingly small \cite{Datta:2008}. 
Quantum discord is originally defined as the misalignment of two classically equivalent expressions for mutual information. Since it was discovered, there has been significant progress toward understanding the nature of quantum discord. However, an algebraic characterization in terms of classical LHV theories similar to the case of quantum entanglement is still missing. Such a characterization would substantially help to clarify the fundamental aspects of quantum discord that contribute to its nontrivial quantum-informational resources.

In this paper, we introduce a special class of LHV representations that serves as a unified framework to characterize both entanglement and discord. In such representations, the linear structure of quantum measurement operators is preserved. It is first identified that LHV models for entangled states, when extant, must be nonlinear. By examining the algebraic structure of the underlying axiomatic probability theory, it is proven that only separable states with zero quantum discord admit fully consistent classical interpretations. 
%Hence, the nature of quantum discord is given a firm footing. It is suggested that the mechanism for the nonclassicality of quantum discord is fundamentally different than that of entanglement. 
The physical significance of linearity, as well as possible generalization of this linear framework to the quasiprobability representation of multipartite quantum states are also discussed. (For simplicity, results in this paper are presented for bipartite systems only. It is straightforward to generalize to multipartite systems.)

\section{Linear LHV representation}
In order to give a more formal discussion of the LHV representations, it is worthwhile to introduce some basics of the axiomatic framework of classical probability theories. We adopt here the measure-theoretic formulation as laid out in the fundamental work of Kolmogorov \cite{Kolmogorov:1956,Athreya:2006}. In this framework, classical probabilities are underlain by a measure space $(\Omega,\mathcal{F},\mu)$, where $\Omega$ is a nonempty set which is called the event space; $\mathcal{F}$ is a set of the subsets of $\Omega$; and $\mu$ is a non-negative set function defined on $\mathcal{F}$. An element $A\in\mathcal{F}$ represents a physically measurable event, whose probability is given by the probability measure $\mu(A)$. The test space $\mathcal{F}$ must contain the empty set and be closed under complementary and unions. 
If these conditions are fulfilled, $\mathcal{F}$ is called a $\sigma$-algebra \footnote{Precisely speaking, a $\sigma$-algebra is closed under countable unions whenever it has an infinite number of elements.}. The $\sigma$-algebra structure, which has been largely overlooked in the study of LHV theories, is an intrinsic property that any classical probability theories must satisfy. As will be shown in the following, it has significant physical consequences. 

%In a classical world, one is able to test the probability that nothing is happening. If the probability of an event $A$ is measurable, then one must be able to ask the probability that $A$ does not happen. Moreover, if another event $B$ is testable, the probability that both $A$ and $B$ are happening must also be testable. These \emph{a priori} reasonings amount to the axiomatic structure of the test space $\mathcal{F}$; that is, $\mathcal{F}$ must contain the empty set and be closed under complementary and unions. If the above conditions are fulfilled,  $\mathcal{F}$ is called a $\sigma$-algebra \footnote{Precisely speaking, a $\sigma$-algebra is closed under countable unions whenever it has an infinite number of elements.}. Probability $\mu$ is required to be normalized, i.e., $\mu(\Omega)=1$. The $\sigma$-algebra structure, which has been largely overlooked in the study of LHV theories, is an intrinsic property that any classical probability theories must satisfy. As will be shown in the following, it has significant physical consequences. 

Since an event $A\in\mathcal{F}$ is a subset of the event space $\Omega$, one may define a corresponding Boolean-valued function $\chi^A_\xi$, the indicator function, to indicate which elements it contains; namely, $\chi^A_\xi=1$ if $\xi\in A$, and $\chi^A_\xi=0$ if $\xi\not\in A$. In this manner, the probability of event A can be re-expressed as an integral: 
\begin{equation}
\mu(A)=\int_\Omega d\mu_\xi \chi^A_\xi.
\end{equation}
Each indicator function represents a deterministic (sharp) measurement of an event. However, in practice there exist indeterministic measurements; i.e., one can measure a combination of different events $\{A_i\}$ with uncertainty specified by the conditional probabilities $\{p_i\}$, where $p_i \geqslant 0$ and $\sum_i p_i \leqslant 1$. In this case, the probability of this indeterministic measurement is given by 
\begin{equation}
\sum_i p_i \mu(A_i)= \sum_i p_i \int_\Omega d\mu_\xi \chi^{A_i}_\xi \equiv \int_\Omega d\mu_\xi f_\xi,
\end{equation}
where $f_\xi=\sum_i p_i \chi^{A_i}_\xi$ is the corresponding response function or measurable function \footnote{Integration over general measurable functions which do not have decompositions into indicator functions can also be rigorously defined \cite{Athreya:2006}. In the present paper, it is sufficient to regard the integration as a linear functional wherever it is used.}; it has the property that $f_\xi\in[0,1], \forall\xi\in\Omega$. The class of all response functions on $\Omega$ is denoted as $\mathcal{T}(\Omega)$.

For composite bipartite systems, an LHV description requires that each local subsystem has its own underlying event space and $\sigma$-algebra. The composed event space is given by Cartesian product $\Omega_a\times\Omega_b\equiv\{(\xi,\eta) : \xi\in\Omega_a, \eta\in\Omega_b\}$, where $a$ and $b$ are the labels for the two subsystems. The joint probability distribution is then represented by a joint probability measure. 
In quantum mechanics, a single measurement event is represented by an effect, which is a positive Hermitian operator that appears in the range of a positive-operator-valued measurement (POVM) \footnote{An effect is a positive Hermitian operator that is bounded above by the identity operator. For more information, refer to, e.g., P. Busch, M. Grabowski, and P. Lahti, Operational Quantum Physics (Springer, Berlin, 1995).}. Denote $\mathcal{E}(\mathcal{H})$ the set of all effects on the Hilbert space $\mathcal{H}$. 
We say that a bipartite quantum state $\rho$ on $\mathcal{H}^a\otimes\mathcal{H}^b$ admits an LHV model if there exists a probability measure $\mu$ defined on a bipartite product measurable space $(\Omega_a\times\Omega_b, \mathcal{F}_a\times\mathcal{F}_b)$, such that for any quantum measurement $M^a\in\mathcal{E}(\mathcal{H}^a)$ and $M^b\in\mathcal{E}(\mathcal{H}^b)$ on subsystems $a$ and $b$, respectively, there exist corresponding measurable functions $f^a(M^a)\in\mathcal{T}(\Omega_a)$ and $f^b(M^b)\in\mathcal{T}(\Omega_b)$, which reproduce the probability of the joint quantum measurement:
\begin{equation}
\label{eq_rep1}
\tr(\rho M^a\otimes M^b)= \int_{\Omega_a\times\Omega_b} d\mu_{\xi,\eta} f^a_\xi(M^a)f^b_\eta(M^b).
\end{equation}
The mappings $f^i: \mathcal{E}(\mathcal{H}^i)\mapsto\mathcal{T}(\Omega_i)$, $i\in\{a,b\}$ are usually not required to be convex linear \footnote{A function $f$ is convex-linear if it is linear on convex combinations, i.e., $f(\sum_i p_i x_i)=\sum_i p_i f(x_i)$, $\forall p_i\in[0,1]$ and $\sum_i p_i=1$}. In fact, all known LHV models for entangled states break the linear structure. When both of the mappings are convex linear, we call the corresponding classical probability representation a linear LHV representation. The physical significance of linearity will be discussed by the end of the present paper. 

\section{Characterizing entanglement}
The following result first establishes a connection between quantum entanglement and such linear representations, which has a nontrivial implication to the framework of quasiprobability representation, as will be outlined in this paper. This result can be viewed as a generalized version of the multipartite extension \cite{Barnum:2005,*Wallach:2000,*Klay:1987,Barnum:2010,*Acin:2010} of the celebrated Gleason's theorem \cite{Gleason:1957}.

\textbf{Lemma 1}. A quantum state of a bipartite system has a linear LHV representation if, and only if, it is separable.

\emph{Proof}. The ``if'' part is trivial. Suppose the state $\rho$ has a separate form $\rho=\sum p_i\rho^a_i\otimes\rho^b_i$; one immediately sees that $\tr(\rho M^a\otimes M^b)=\sum p_i \tr(\rho^a_iM^a)\tr(\rho^b_iM^b)$ \cite{Werner:1989}. To prove the ``only if'' part, assume that there exists a joint probability measure $\mu_{\xi,\eta}$ on a product space $(\Omega_a\times\Omega_b,\mathcal{F}_a\times\mathcal{F}_b)$, such that any product measurement on the quantum state $\rho$ can be represented as a joint classical (indeterministic) measurement given by Eq. (\ref{eq_rep1}).
By definition, $f^a_\xi\in[0,1]$ is a convex-linear functional. Using the same technique described in Ref. \cite{Busch:2003}, it can be uniquely extended to a linear map on the space of all bounded Hermitian operators. As a consequence of the Riesz representation theorem, there is a unique positive-Hermitian operator-valued functional $F^a_\xi$ on the Hilbert space of subsystem $a$, such that
\begin{equation}
\label{eq_repf}
f^a_\xi(M^a)= \langle F^a_\xi,M^a \rangle,
\end{equation}
where $\langle A, B\rangle = \tr (A^{\dagger}B)$ is the Hilbert-Schmidt inner product. The same procedure also applies to $f^b_\eta$, which is then represented by a unique operator-valued functional $F^b_\eta$ on system $b$. For identity operators $I^a$ and $I^b$, $\int\ d\mu_{\xi,\eta} \  f^a_{\xi}(I^a) f^b_{\eta}(I^b) = \tr(\rho I^a\otimes I^b)=1$; therefore $f^a_{\xi}(I^a) =f^b_{\eta}(I^b)=1$. Consequently, $F^a_\xi$ and $F^b_\eta$ are both unit-trace operators, which are legitimate as quantum states.
Equation (\ref{eq_rep1}) can be expressed as
\begin{equation*}
\label{rq_rep2}
\begin{aligned}
\tr(\rho M^a\otimes M^b)=&\int d\mu_{\xi,\eta} \langle F^a_\xi,M^a \rangle \langle F^b_\eta,M^b\rangle\\
=&\tr\int d\mu_{\xi,\eta}F^a_\xi\otimes F^b_\eta M^a\otimes M^b.
\end{aligned}
\end{equation*}
The above equation is valid for any effect, which implies that 
\begin{equation}
\label{eq_rho}
\rho=\int_{\Omega_a\times\Omega_b} d\mu_{\xi,\eta}F^a_\xi\otimes F^b_\eta.
\end{equation}
The separability of $\rho$ can be seen from the fact that it is non-negative for any entanglement witness.\hfill$\Box$

\section{Characterizing discord}
The above result shows that all separable states admit LHV interpretations with every local quantum measurement being represented by a corresponding response function. However, the $\sigma$-algebra, which supports these response functions, has not been examined yet. Given a LHV representation, it is possible that there are fundamental measurements, i.e., measurements that correspond to indicator functions, which do not have counterparts in the quantum description. This is not acceptable in a consistent classical probability theory, where the test space represents fundamentally measurable events. The fact that classical mechanical systems are described by variables that are ``not hidden, but in principle measurable" has been noticed and pointed out by Bennett \emph{et al.} \cite{Bennett:1999}. We call an LHV representation \emph{tight} when each measurable subset in its $\sigma$-algebra is physically testable; i.e., for each indicator function there exists a corresponding quantum measurement that corresponds to it. To phrase this mathematically, for tight representations, the mapping $\mathcal{E}(\mathcal{H})\mapsto\mathcal{T}(\Omega)$ covers the set of all indicator functions. Intriguingly, not all separable states admit such fully consistent classical descriptions in the sense of tight representations. The following theorem establishes a connection between tight LHV representations and zero quantum discord. 

\textbf{Theorem 2}. A quantum state of a bipartite system has a \emph{tight} linear LHV representation if, and only if, it has a null quantum discord.

Before presenting the proof, it is worthwhile to briefly discuss the notion of quantum discord \cite{Ollivier:2001,Henderson:2001}. Correlations between two quantum systems are captured by the mutual information
\begin{equation}
I(a:b)\equiv S(\rho^a)+S(\rho^b)-S(\rho^{ab}),
\end{equation}
where $S(\rho)=-\tr(\rho\log\rho)$ is the von Neumann entropy. Another way to qualify the mutual information is to first perform a POVM on one system, e.g., $\{M^a_i\}$ on system $a$, which collapses system $b$ to a set of conditional states $\{\rho^b_i\}$ with the corresponding probabilities $\{p_i\}$. This induces the second measure of mutual information
\begin{equation}
J(b|\{M^a_i\})\equiv S(\rho^b)- \sum_i p_i S(\rho^b_i),
\end{equation}
the last term of which is the conditional entropy.
The classical counterparts of the above two expressions are equivalent. However, as was identified in Ref. \cite{Ollivier:2001,Henderson:2001}, they are generally not equal in the quantum case. The second expression is measurement-dependent and no greater than the former one. Their minimum difference with respect to all POVMs is defined as the quantum discord:
\begin{equation}
D(b|a)\equiv I(a:b)-\max_{\{M^a_i\}} J(b|\{M^a_i\}),
\end{equation}
which is asymmetric under the exchange a $\leftrightarrow$ b. We say a state $\rho$ has a null quantum discord when both $D(b|a)$ and $D(a|b)$ are zero. It has been proven \cite{Hayashi:2006,Datta:2010} that a necessary and sufficient condition for $D(b|a)=0$ is the existence of decomposition
\begin{equation}
\label{eq_condition}
\rho=\sum_i p_i \Pi^a_i\otimes\rho^b_i,
\end{equation}
where $p_i\geqslant0$, $\{\Pi^a_i\}$ is a complete set of rank-1 projection operators on system a, and $\rho^b_i$ are density operators of system b. The equivalent condition for $D(a|b)=0$ can be similarly induced by decomposing $\rho$ with rank-1 projection operators on system b. State (\ref{eq_condition}) is called the classical-quantum state, which is a fundamental building for measures of quantum correlations\cite{Modi:2010}. Its nonclassical nature can be viewed from various perspective\cite{Oppenheim:2002,Li:2008,Madhok:2011,Perinotti:2012,roga:2016}. The result of theorem 2 provides an algebraic characterization by directly comparing with the classical probability theory.

\emph{Proof of Theorem 2}. We first prove the ``only if" part. Suppose a state admits a classical probability interpretation given by Eq. (\ref{eq_rep1}). Consider first the space ($\Omega_a,\mathcal{F}_a$). 
For any two elements $\xi,\xi'\in \Omega_a$, if there exists a measurable subset $\Delta\in\mathcal{F}_a$ such that it does not contain $\xi$ and $\xi'$ simultaneously (without losing generality we assume $\xi\in\Delta$ and $\xi'\not\in\Delta$), one can find a corresponding indicator function $\chi^\Delta_\xi$ defined on $\Omega_a$ such that $\chi^\Delta_\xi=1$ and $\chi^\Delta_{\xi'}=0$. Since $\Delta$ is physically measurable, according to Lemma 1, there exists a corresponding effect $M^a_\Delta$ satisfying Eq. (\ref{eq_repf}), which leads to the constraints
\begin{equation*}
\langle F^a_\xi,M^a_\Delta \rangle=\chi^\Delta_\xi=1\  \textrm{and}\  \langle F^a_{\xi'},M^a_\Delta \rangle=\chi^\Delta_{\xi'}=0.
\end{equation*}
The first of the above equations implies that the support of $F^a_\xi$ is contained in the 1-eigenspace of effect $M^a_\Delta$, while the second implies that the support of $F^a_{\xi'}$ lies in the kernel of $M^a_\Delta$. Therefore, the supports of $F^a_\xi$ and $F^a_{\xi'}$ are orthogonal subspaces of $\mathcal{H}^a$. Conversely, if $\xi$ and $\xi'$ always appear together in any measurable subset in $\mathcal{F}_a$, then any measurable function, as a linear combination of indicator functions, would produce the same value on $\xi$ and $\xi'$. According to Eq. (\ref{eq_repf}), $\langle F^a_\xi,M^a\rangle=\langle F^a_{\xi'},M^a\rangle$ for any effect $M^a$, thus $F^a_\xi=F^a_{\xi'}$. Hence, any two operators in $\{F^a_\xi\}$ are either equal or orthogonal. Since $\mathcal{H}^a$ is finite dimensional, there is a finite number of distinct orthogonal states, which we label as $\{F^a_i\}$. One can draw the same conclusion and get a finite set of orthogonal states $\{F^b_j\}$ for system $b$. 

The tensor product operators $\{F^a_i\otimes F^b_j\}$ span a subspace of $\mathbb{H}(\mathcal{H}^a\otimes\mathcal{H}^b)$, where $\mathbb{H}(\mathcal{H})$ denotes the real Hilbert space of Hermitian operators on $\mathcal{H}$, equipped with the Hilbert-Schmidt inner product. Since $\rho$ lies in this subspace, it can be expanded in terms of the orthogonal basis $\{F^a_i\otimes F^b_j\}$ with corresponding expansion coefficients
\begin{equation*}
\begin{aligned}
p_{ij}=\langle\rho,F^a_i\otimes F^b_j\rangle =\int_{\Omega_a\times\Omega_b} d\mu_{\xi,\eta} \langle F^a_\xi,F^a_i\rangle \langle F^b_\eta,F^b_j\rangle\geqslant0,
\end{aligned}
\end{equation*}
where the Boolean-valued functions $\langle F^a_\xi,F^a_i\rangle$ and $\langle F^b_\eta,F^b_j\rangle$ are indicator functions.
The composite state (\ref{eq_rho}{}) is then reduced to
\begin{equation}
\label{eq_discord}
\rho=\sum_{i,j}p_{ij}F^a_i\otimes F^b_j.
\end{equation}
$F^a_i$ and $F^b_j$ can be further decomposed into rank-1 projection operators. According to the sufficient condition introduced in Eq. (\ref{eq_condition}), it is concluded that $\rho$ has a null quantum discord.

On the other hand, for a null quantum discordant state with decomposition $\rho=\sum_{i,j}p_{ij}\Pi^a_i\otimes\Pi^b_j$, one can define an event space $\Omega_a\times\Omega_b\equiv\{(\xi^a_i,\xi^b_j)\}$ labeled by the same index $i$ and $j$ as in the decomposition, together with a joint probability measure $p_{ij}$. The $\sigma$-algebra $\mathcal{F}_a\times\mathcal{F}_b$ is defined as the power set, the set of all subsets, of the event space. It is easy to see that every element $(\xi^a_i,\xi^b_j)$ has a corresponding quantum mechanical counterpart $\Pi^a_i\otimes\Pi^b_j$. Hence, every measurable subset in the test space is physically realizable. This proves that the classical representation is tight. \hfill$\Box$

The above theorem can be generalized to characterize states with only one zero discord, say, $D(b|a)=0$ and $D(a|b)\not=0$. It is not hard to show that such states have LHV representations that are tight only on system $a$.

\section{Discussion}
As an effective framework, the linear LHV representation developed above is powerful and of interest on its own. We now argue that the requirement of linearity has profound physical significance. Quantum measurements are represented by effects which are linear operators. The trace formula guarantees that probabilities are linear on measurements as well. This means for effects with linear decomposition, for instance, $M=M_1+M_2$, the corresponding probabilities satisfy $p(M)=p(M_1)+p(M_2)$ independent of the states they are acting on. This is an intrinsic property of the quantum measurements themselves. In a non-linear LHV model which maps these effects to measurable functions such that $f(M)\not=f(M_1)+f(M_2)$, the linear condition for the corresponding probabilities is no longer generally true. In other words, the complete information of quantum measurements is not fully encoded by a non-linear mapping. %Such an LHV description predicts inconsistent behaviors of the measurement effects compared with the quantum ones.

One might argue that the objective is not to endeavor to have a complete description of quantum measurements. Instead, the measurable functions are only designed to work for the probability measure $\mu_\rho$ of the target state $\rho$ itself, over which $\int d\mu_\rho f(M)=\int d\mu_\rho f(M_1)+\int d\mu_\rho f(M_1)$. The inconsistency occurs for other probability measures that are irrelevant to the LHV model of $\rho$. The mapping $\mathcal{E}(\mathcal{H}) \mapsto \mathcal{T}(\Omega)$ nevertheless produces all the statistics of local quantum measurements on $\rho$. However, as was suggested by various protocols \cite{Popescu:1995,Gisin:1996,Peres:1996,Masanes:2008,Buscemi:2012} that extract the hidden nonlocality, a quantum state usually reveals its hidden features when jointly considered with other states. Instead of studying the target quantum state alone, one ought to examine its behavior among a set of quantum states as a whole. This perspective leads to a grand picture of representing both the set of all quantum measurements and the set of all the quantum states of the system. Needless to say, it is well known that quantum mechanics has no classical representations. In order to achieve this picture, the probability measures are allowed to take negative values. Rigorously speaking, each quantum state is mapped to a signed measure (quasiprobability) on the event space $\Omega$. This unified framework, called quasiprobability representation, has been shown \cite{Ferrie:2009,Ferrie:2011} to always exist. Wigner quasiprobability distribution \cite{Wigner:1932} is the most well known example and has wide applications in various fields in physics. Here we emphasize that in such framework, linearity, instead of being an additional physical assumption, is a direct mathematical consequence (see Lemma 2 in Ref. \cite{Ferrie:2009} for a proof). However, as was pointed out \cite{Ferrie:2011}, the quasiprobability representation of single quantum systems does not induce any characterization of quantum states at all: For any quantum state $\rho$ of a single system, one can always find a suitable representation, in which $\rho$ is mapped to a positive probability measure (there must be, of course, other states that are mapped to negative ones). Lemma 1 of the present work shows that the situation changes dramatically when the local structures of compound systems are introduced: Entangled states are those that can not be mapped to positive probability measures in any quasiprobability representations. We expect that the introduction of quasiprobability measures in LHV theories could inspire new insights in studies of the fine structures of entanglement, such as the long-standing bound entanglement problem \cite{Horodecki:1998,Clarisse:2006}.

To conclude, this paper developed a framework of linear LHV representations of quantum states of composite systems, in terms of algebraic probability theory, which allows us to have a unified characterization of both entanglement and quantum discord. Entangled states are characterized as those states that do not admit linear LHV representations. This provides new insights in understanding the relation between entanglement and nonlocality. By examining the algebraic structure of the underlying probability theory, it is shown that fully consistent classical interpretations are only accessible to zero discordant states. A new mechanism that might contribute to the nonclassicality of quantum discord is thus identified. Potential generalization of this framework to the quasi-probability representation of multipartite quantum states and applications to the study of entanglement theory were also discussed.

%\begin{acknowledgments}
This work was supported in part by NSF grant PHY-1607180. The author is indebted to Professor Chris H. Greene for his kind support; The anonymous referee is also greatly acknowledged, especially for his/her suggestions that further simplified the proof of theorem 2.
%\end{acknowledgments}

\bibliography{reference}
\end{document}